# PHYSICS OF A SUPERFLUID SOLID

P W Anderson, Princeton University

ABSTRACT

The question of whether a Bose solid has a superfluid fraction is discussed. We confirm and expand the remark of Chester that quantum fluctuations make this inevitable, roughly estimate its magnitude, and show that true rigid rotation depends on the generation of vortex defects. We also suggest a model wave-function for the superfluid solid.

_________________________________  ____________

Moses Chan and coworkers[1] have recently found experimental evidence that below about 200 millidegrees K solid helium has a non-rotating (superfluid) component amounting to roughly 1% of its moment of inertia. Many experimental checks failed to find any evidence of defects or grain boundary phenomena in the samples; solid He normally grows in very defect-free crystals. In addition, it seems to us that the fact that He in porous substrates does not behave appreciably differently is strong evidence against any defect explanation. More recently, similar phenomena have been observed in H2.

A number of theoretical papers have suggested that exchange effects might lead to superfluidity in solids, notably those by Chester[2] and Leggett[3]. Chester's proposal was in turn based on a theorem of Reatto[4]. The physics here proposed does not differ in principle from their ideas; the intent here is to flesh out the phenomenology to show that it is compatible with the experiments, and that Leggett's limits on the magnitude of Chester's effect could have been too stringent. We also introduce a model wave function which may be useful in understanding the phenomenology.

Theories of quantum solids come in many varieties. The original work of Chester and coworkers was based on the Bijl-Jastrow wave function as used in the helium context by MacMillan[5]. The central idea here is that the trial wave function, since it is everywhere positive, can be thought of as a classical Boltzmann function for atoms interacting via a two-body potential:

$$\psi(r_1, r_2, \cdots, r_N) = Q_N^{-1/2}(\beta u) \exp[-\frac{1}{2} \sum_{i,j}^{N} \beta u(r_i - r_j)] \quad [1]$$

Here Q is the normalization constant (also equal to the classical partition function) and u(r) is a two-body fictitious potential chosen to optimize the calculated energy. It is very large at r=0 and falls off to zero rapidly with distance. Its longest-range part is determined by the phonon velocities but does not much concern us here.

The wave function [1] seems to be a remarkably accurate description of the properties of liquid He as shown in ref [5] for instance; that it also represents the solid at higher densities is evident from the fact that the classical system will surely solidify for sufficiently large βu; Macmillan produced a fairly good estimate of some solid properties.

In the limit β→∞ [1] becomes a perfect solid with each r near a potential minimum at a lattice site. A wave function with this property does not have ODLRO according to the criteria in refs [2] and [4], because the variables $r_2 \cdots r_N$ absolutely fix the position of $r_1$. It cannot be superfluid in any ordinary sense.

But the important fact about [1] is that $\beta \neq \infty$. The equivalent classical solid is at a finite temperature, so there is a finite density both of interstitials and of vacancies. This density will be proportional to

$$n_V \propto \exp{-\beta E_V}, \quad n_I \propto \exp{-\beta E_I},$$

where the E's are the finite energies of a vacancy or an interstitial in the equivalent classical crystal.
The superfluid density, therefore, which can be shown by the methods of reference [4] to be roughly equal to the density of vacancies or interstitials (and I suspect vacancies will tend to predominate) will always be finite; the Bose solid is never a "true" solid in the sense of ref [6]. The question then is not whether or not the solid is superfluid but to estimate its properties.

Much of the complication of the theories of the quantum solid comes from the attempt to take into account the large zero-point amplitude of the phonons, which carries them outside of the linear range of the interatomic potential and makes numerical computation very difficult. Here however we are not very interested in numerical accuracy, nor is it likely we could achieve it. For our purposes it may be adequate to discuss a model wave function for the solid based on the simple Hartree-Fock theory sketched in my book[6]. That theory can be thought of as a model description of the physical fact that there is a crystalline lattice and that there is a helium atom occupying the overwhelming preponderance of its sites.

The wave function proposed in ref 6 is

$$\Psi = (\prod_i c_i^*)\Psi_{vac} \quad [2]$$

where the c's are a set of localized self-consistent boson operators referring to orbitals localized at the sites i of a lattice:

$$c_i^* = \int d^3r \phi_i(r)\psi^*(r) \quad [3]$$

$\psi^*(r)$ is the boson creation operator. I showed in reference [6] that the wave functions $\phi_i(r)$ satisfy a self-consistency equation obtained as the Hartree-Fock equation for hole (vacancy) excitations (which in the Bose case is not the same as that for particle excitations.) If the potential is sufficiently repulsive, this equation can have a self-consistent localized potential well because, in the boson case, for holes the exchange term adds to the self-consistent potential rather than compensating it as for fermions. For particles, on the other hand, the effective potential is perfectly periodic and cannot have a bound state—hence there is an energy gap separating the particle (interstitial) states from the hole (vacancy) ones. Hartree-Fock theory thus nicely expresses the fact that an added particle sees the repulsive potential of all occupied sites, while the atom occupying the site sees only its neighbors. The solid is stable at a density fixed (hypothetically) by optimizing the energy. The localized wave functions $\phi(r)$ are by no means orthogonal to each other, since unlike the Fermi case the potential differs for each i.  The potential well arises as a consequence of the fact that the particle cannot interact with itself: it is there because each site contains exactly one particle which is repelled by its neighbors but not by its own potential.  The hole-particle gap occurs when there is exactly one particle per site.  This Hartree-Fock Bose solid is what I called a "true" solid, equivalent to an insulator in the electronic analog, in that there is a downward cusp in the energy as a function of occupancy of the sites at exactly one—or an integer number—per site.  It is also locally gauge invariant and cannot be superfluid.

But the Hartree-Fock wave function above, which we imagine to have been calculated using a pseudopotential rather than the true hard core potential, cannot be correct. The true wave function must contain a finite density of vacancies as explained above; and also the overlap of the $\phi$'s is unphysical because two particles cannot have r-r'$\cong$0, because of the hard core of the potential.

These two deficiencies (which are closely related to each other) can be remedied by  a "BCS-like" modification of the wave function [2] and simultaneously of the orbitals [3]:

$$\Psi = \prod_{i}^{N'} (g + e^{i\theta_i} c_i'^*) \Psi_{VAC}; \quad N' = N(1 + g^2) \quad [4]$$

$g^2$ is the average density of vacancies.
The orbitals are to be artificially orthonormalized:

$$c_i'^* = \int \phi_i'(r)\psi^*(r)dr \text{, where } \int \phi_i'(r)\phi_j'(r)dr = \delta_i^j \quad [5].$$

As a consequence there is no double occupancy of an orbital—a kind of "Gutzwiller" constraint.
It is well-known that to compensate for such an orthogonalization we must introduce a kinetic energy matrix element

$$t_{ij} = S_{ij} E_V \text{ where } S_{ij} \text{ is the overlap of i and j}, \quad [6]$$

and E is the vacancy energy. This mends the deficiency in ref [6] that the vacancy had no kinetic energy. The wave function [4] allows us to define a local phase θ, which is the phase of φ relative to g, and is coarse-grained on the scale of the atoms. The energy calculated using the wave function [4] contains a term

$$\sum_{i<j} t_{ij} g^2 \cos(\theta_i - \theta_j) \quad [7]$$

which implies that there is a supercurrent

$$\mathcal{J} = \text{const} \times \nabla \vartheta \qquad [8]$$

which need not express Galilean invariance because the lattice serves as a preferred reference frame.

Leggett in ref [3] gave a rough estimate of the magnitude of "const" in terms of the exchange energy in solid He3. I think this estimate can be too low, because of the following argument.

In reference [4] I showed that the potential well which binds the localized state at site i is given by

$$\Delta V_i(r) = \int d^3 r' V(r-r') \Delta \rho_i(r')$$

plus a nonlocal "Fock" term of slightly more complicated form but similar magnitude and sign. $\Delta\rho$ is the missing density at site i

$$\Delta \rho_i = -[|\phi_i|^2 + \sum_j S_{ij} \phi_i(r)\phi_j(r)]/(1+\sum_j |S|_{ij}^2).$$

It is important to note that the Fock term is of similar magnitude and of the *same* sign.

The effects of exchange and of the direct overlap are opposite for He3 spins. Exchange only operates for parallel spins, and is ferromagnetic in sign; while the nonorthogonality effect is antiferromagnetic, preferring antiparallel near neighbors. It seems likely that the two nearly cancel in He3, leading to an anomalously low amplitude of spin exchange. This is supported by the observation that loop exchange has a surprisingly large relative effect in that case.

We can define the constant in [7] or [8] in terms of a heavy "effective mass" M by

$$v = \frac{\hbar}{M}\nabla\phi \qquad [5]$$

The energy scale of [7] is then of order $\hbar^2/Ma^2$.

It is, however, essential that when the lattice of sites and its superfluid fraction (which is what [7] is) are set in uniform motion, the whole mass moves with them. Thus when there is a constant gradient of the phase, [5] and the "normal" fraction J of the current, that which is carried along with the sites, must add up to the total mass of helium—the normal fraction is less than unity.

Because of [4] the supercurrent has a constraint:

$$\nabla \times \mathcal{J} = 0 \qquad [9]$$

Of course, if [9] is satisfied everywhere, then the current can only be 0, or at most a constant; but if we allow for a line defect where we can make Ψ=0, that is we break a one-dimensional manifold of our bonds, we can satisfy them with a vortex flow with v∝1/r.

Now if we try a rotation experiment, the rigid rotation of the site lattice does not obey $\nabla \times v = 0$, so that in the absence of a vortex singularity the superfluid fraction cannot participate in the rotation at absolute zero. As we raise the rotation velocity, at low temperatures vortices will be drawn into the sample in order to mimic rigid rotation a la Onsager-Feynman. There will be a critical angular velocity, analogous to $H_{c1}$ of a type II superconductor, when the rotation energy is first equal to the cost of vortices. I estimate that because the vorticity unit of the superflow is small($\propto 1/M$), this is quite small, of the order of one quantum of ordinary vorticity, and comparable to the observed threshold. (The smallness of 1/M cancels the large logarithm $\ln(R^2/a^2)$ in the vortex energy.)

Beyond the threshold the superfluid fraction will drop steeply (logarithmically). As the temperature rises, thermally activated vortices will gradually appear and allow the solid to equilibrate in the rigidly rotating state; the transition region will be dissipative like the vortex liquid of ordinary superconductors. The genuinely rigid, solid phase is a liquid of free vortices, in which forces will be accommodated by vortex motion or, equivalently, free phase slippage, rather than by superleak currents. Isotopic impurities will nucleate vortices and thus will tend to destroy the superfluidity and restore rigidity.

It seems that this scenario is compatible with all the observations so far, although of course it badly needs a more quantitative approach. One interesting additional experiment has been suggested to me[7]: The supercurrent will couple very differently to longitudinal phonons, since for these ∇×J≈0, relative to transverse ones. The longitudinal phonons may partially take on particle character, though the effect may be very small; while the transverse phonons will have a velocity change similar to that of the moment of inertia.

It is significant that the familiar rigid solid turns out very generally not to be rigid, though normally only near absolute zero; rigidity seems to be actually an emergent phenomenon of the classical limit..

I would like to acknowledge extensive discussions with M H W Chan and with W F Brinkman, without which this paper would not have been written.